# The collaboration behavior of top scientists

Giovanni Abramo

*Laboratory for Studies in Research Evaluation*
*at the Institute for System Analysis and Computer Science (IASI-CNR)*
*National Research Council of Italy*
ADDRESS: Istituto di Analisi dei Sistemi e Informatica, Consiglio Nazionale delle Ricerche, Via dei Taurini 19, 00185 Roma - ITALY
giovanni.abramo@uniroma2.it

Ciriaco Andrea D'Angelo

*University of Rome "Tor Vergata" - Italy and*
*Laboratory for Studies in Research Evaluation (IASI-CNR)*
ADDRESS: Dipartimento di Ingegneria dell'Impresa, Università degli Studi di Roma "Tor Vergata", Via del Politecnico 1, 00133 Roma - ITALY
dangelo@dii.uniroma2.it

Flavia Di Costa

*Research Value s.r.l.*
ADDRESS: Research Value, Via Michelangelo Tilli 39, 00156 Roma- ITALY
flavia.dicosta@gmail.com

**Abstract**
The intention of this work is to analyze top scientists' collaboration behavior at the "international", "domestic extramural" and "intramural" levels, and compare it to that of their lesser performing colleagues. The field of observation consists of the entire faculty of the Italian academic system, and so the coauthorship of scientific publications by over 12,000 professors. The broader aim is to improve understanding of the causal nexus between research collaboration and performance. The analysis is thus longitudinal, over two successive five-year periods. Results show a strong increase in the propensity to collaborate at domestic level (both extramural and intramural), however this is less for scientists who remain or become top, than it is for their lower-performing colleagues. In contrast, the increase in international collaboration behavior is greater for scientists who become or remain top than it is for their peers. The increase in productivity by those who acquire top scientist status is due precisely to the greater average impact of the publications achieved in collaboration with foreign colleagues.

# 1. Introduction

The scientific world is experiencing remarkable growth in collaboration for the purposes of research. This reality has been confirmed, for decades, by the analysis of coauthorship (Melin & Persson, 1996), indicating that the share of single-authored publications is in constant decline (Abt, 2007; Uddin, Hossain, Abbasi, & Rasmussen, 2012), and that there is a corresponding increase in average of number of authors per publication (Wuchty, Jones, & Uzzi, 2007; Gazni, Sugimoto, & Didegah, 2012; Larivière, Gingras, Sugimoto, & Tsou, 2015). The collaboration phenomenon is particularly evident in large-scale scientific research, or 'big science', where participation at the knowledge frontier requires major infrastructure, large coordinated groups of scientists and technicians, backed by substantial funding over periods of many years. For academics pursuing a scientific career, it becomes ever more important to develop collaborations with colleagues in their own and other universities and with other domestic and international institutions. This enables participation in broader research projects, access to funding, as well as the very important aspect of improvement in personal competences, with positive effects on the individual's quantity and quality of publications. Many countries have implemented policies incentivizing scientists' capacities to activate and manage effective collaborations. The pursuit of collaborations is now recognized as a rewarding personal strategy, particularly for career advancement.

The benefits from research collaboration are observed at levels beyond the individual, extending to various aggregations, to institutions as a whole (universities, research agencies), and entire countries (national research systems). However the effects and costs to be considered will vary according to the type of collaboration: domestic versus international, intramural versus extramural, intrasectoral versus intersectoral. Collaboration behaviors at the individual level can also vary in intensity in relation to: i) contextual factors, most obviously the scientific research discipline (Abramo, D'Angelo & Murgia, 2013a; Gazni, Sugimoto, & Didegah, 2012; Yoshikane & Kageura, 2004), but also the sectoral diversity of the research projects scientists participate in: and ii) personal factors such as gender, age, academic rank (Kyvik & Olsen, 2008; Bozeman & Gaughan, 2011; Gaughan & Bozeman, 2016). There are also the many matters of social conventions, such as those concerning the assignment of authorship and credits for publications (Katz & Martin 1997; Cronin, 2001).

The intention of this work is verifying whether the collaboration behavior of top scientists (TSs) is substantially different from that of their colleagues, and if these differences then vary across disciplines and fields. The broader aim is to improve understanding of the causal nexus between research collaboration and performance. For reasons which will be explained in Section 3, the Italian academic system provides a useful and valid field of observation. Within the population of all Italian university faculty, a bibliometric approach is used for identification of the TSs and for the measurement of their collaboration behavior.

The next section examines the existing literature, attempting to understand the links between research collaboration, performance, and other personal and organizational variables. Section 3 describes the dataset and the methodology used, to further this understanding. Section 4 presents the empirical results, while the final section proposes additional avenues of investigation and offers some policy recommendations.

## 2. Literature review

The term 'scientific collaboration' presents difficulties, in both definition and eventual measurement. The concept is that of a social convention developed between scientists at various levels of the research system, however the modes of this convention are not always easy to distinguish. Observers consider that over time, academic research has become ever more a social phenomenon (Leahey, 2016; Powell, White, Koput, & Owen-Smith, 2005; Rawlings & McFarland, 2011). This leads to its consideration as a social activity carried out within institutional contexts, rather than as a purely rational strategy to maximize productivity (Bozeman, Dietz, & Gaughan, 2001). According to Bozeman and Boardman (2014), research collaboration involves "social processes whereby human beings pool their experience, knowledge and social skills with the objective of producing new knowledge." Borrowing literature from social capital, Jha and Welch (2010) investigated the extent to which multifaceted collaboration is attributable to the particular relational aspects of the individuals' networks. In this regard, a recent study by Zhang, Bu, Ding and Xu (2018) analyzed scientific collaboration as the effect of homophily, transitivity, and preferential attachment of individual scientists.

In examining the main determinants of social capital, the literature has concentrated above all on the role of academic rank. The greater responsibilities and related resources for senior academics mean that they tend to develop collaboration networks that are broader (Bozeman & Gaughan, 2011), more cosmopolitan (Bozeman & Corley, 2004), better consolidated and more highly productive (Martín-Sempere, Garzón-García, & Rey-Rocha, 2008). Many collaborations involving higher-ranked academics arise from the need to access resources, such as laboratories, equipment, administrative and support personnel. For younger researchers, the failure to involve senior academics would exclude them completely from obtaining such resources. Indeed, it is natural that mentorship relationships generally require full or associate professors in the mentor role and assistant professors in the role of "mentee". In this manner, the lower ranking professors gain advantage from accessing the greater experience and social capital of the full professors. Younger academics seek and are pushed to collaborate, not only to overcome their difficulties in accessing resources, but also to demonstrate their capacities in activating and managing collaborations, considered essential to career progress.

The activation and management of collaborations implies costs that can vary significantly with the age and rank of the relevant academics. Such inequalities are sometimes due to the unequal division of duties in the collaboration, at greater expense to researchers with less power. However, more frequently the differences in tasks is because of the different levels of experience of full, associate and assistant professors (Lee & Bozeman, 2005). According to Abramo, D'Angelo and Murgia (2014), young and lower ranking academics seem to enter into collaborations primarily at the intramural level. Greater numbers of extramural collaborations are observed for full professors, partly due to the opportunities that such faculty have for participation in governance activities, which then permit them to activate links with colleagues from other universities, particularly at the domestic level (van Rijnsoever, Hessels & Vandeberg, 2008). In fact it has been observed that collaborations involving higher numbers of organizations feature a strong presence of senior academics (Hinnant et al., 2012). When it comes to intensity of collaboration at the international level, full professors then play a dominant role with respect to associate and assistant professors (Abramo et al., 2014; Frehill, Vlaicu, & Zippel, 2010). Differences can also be seen in the modes of activating these international collaborations: for full and associate professors the mechanism is more typically through conferences, while for assistant professors the path is through previous participation in PhD programs abroad (Melkers & Kiopa, 2010). The smaller number of international collaborations for assistant professors would in part be explained by the fact that this type of collaboration is not particularly important for progressing to higher ranks (Arthur, Patton, & Giancarlo, 2007), with exceptions in certain disciplines, such as Physics (Ackers, 2005, 2004). Obviously, junior and lower-ranking professors must first establish a solid academic record and reputation, including through domestic collaboration, and then build further through participation in international projects and networks.

Apart from this discussion of academic rank, gender has been identified as a personal factor that influences research collaboration behavior. Female scientists face with multiple challenges. On the one hand they have more difficulty in obtaining the funds necessary to conduct their research, and on the other their motivation is not strengthened by adequate social support. It has been observed that, as a result, female academics tend to develop more formal collaborations (Sonnert & Holton, 1995), and networks of contacts that are both less cosmopolitan (Bozeman & Corley, 2004) and less prestigious (Fuchs, Stebut, & Allmendinger, 2001; Long, 1990). This translates into a gap in the social capital of women academics (Rhoten & Pfirman, 2007), as has often been reported in the mainstream literature. This gap could in turn be a partial or even the full explanation for the lesser productivity of female researchers, documented in studies of different disciplines and in various nations (Abramo, D'Angelo, & Caprasecca, 2009; Larivière, Vignola-Gagné, Villeneuve, Gelinas, & Gingras, 2011; Mauleón & Bordons, 2006).

The link between research collaboration and performance seems at this point to be thoroughly accepted in the literature (Katz & Hicks, 1997; Lee & Bozeman, 2005; Liao, 2011; Carillo, Papagni, & Sapio, 2013; Aldieri, Kotsemir, & Vinci, 2017). However it remains that the causal nexus between these two has not been clarified. One of the motivations prompting individual scientists to collaborate is doubtless this very desire to increase scientific production, yet on the other hand, it is also evident that those who are already highly productive researchers are a favored prospect for scientific collaborations. Abramo, D'Angelo and Murgia (2017) show that research productivity positively influences collaborations at intramural and international level, thanks to the "attraction" exercised by the most productive scientists, and by their greater ability in effectively managing collaborations. But looking from the other side of this nexus (in the same work, and contrary to previous literature), they find that only collaborations at domestic level have a positive impact on research productivity. This result could be explained by the higher costs related to international collaborations, which seem to prevail over the benefits of this form of collaboration.

In this work we intend to explore the issues of the collaboration/performance nexus more deeply, with our "point of entry" being TSs, meaning those scholars who are most productive in their respective fields. Our starting research question is whether their propensity to collaborate is different from that of their peers. In the literature, the only contribution published on this theme seems to be that of Abramo, D'Angelo and Solazzi (2011), according to which researchers with top performance over national colleagues are also those who establish more foreign collaborations (the reverse is not always observed). Their conclusions are based on a cross-sectional analysis, however, they do not inquire into the causal nexus between the collaboration behaviors and the performance of these top scientists. The current paper examines such nexus, through a longitudinal analysis. The methodological details are explained in the next section.

## 3. Data and method

In this section, we present i) the field-classification system of the Italian faculty, object of the analysis; ii) the bibliometric indicator of performance used to identify TSs; and iii) the taxonomy of collaboration behavior and relevant indicators.

A fundamental requirement of this study is the identification of TSs. Given that the intensity of publication varies across fields (D'Angelo & Abramo, 2015), there is a risk of comparing apples to oranges (Abramo, Cicero & D'Angelo (2013) which would cause significant sectoral distortions in performance rankings (Abramo & D'Angelo, 2007). For this, it is necessary to classify the population under observation into rather homogeneous fields of research. This will also allow us also to investigate whether the collaboration behavior of TSs varies across fields. To the best of our knowledge, with the exception of Norway, no country other than Italy has a database of all academics classified by their research field. Furthermore, the Norwegian classification is much coarser than the Italian one, making the Italian case the one best suited to this kind of analysis.

In Italy each professor is classified in one and only one research field named "scientific disciplinary sector" (SDS, 370 in all).[1] SDSs are grouped into disciplines named "university disciplinary areas" (UDAs, 14 in all). The source for data on the faculty at each university is the database maintained by the Ministry of Education, Universities and Research (MIUR),[2] which indexes the name, gender, academic rank, field (SDS/UDA), and institutional affiliation of all professors in Italian universities, as recorded at the close of each year.

The dataset used for the analyses is a subset of the whole population, and is made up of professors who satisfy the following two conditions in the period 2001-2010: i) they are permanently on staff over the whole period, at the same university and SSD; and ii) they have produced at least one authored publication indexed in WoS. Citations are counted as of 30/06/2017, which makes the citation window long enough in time to ensure an acceptable predictive power concerning long-term impact of publications (Abramo, Cicero, & D'Angelo, 2011).

Because the bibliometric repositories' coverage of research output in arts and humanities in unsatisfactory (Hicks, 1999; Archambault, Vignola-Gagné, Côté, Larivière, & Gingras, 2006), to ensure robustness of the bibliometric approach, our analysis deals only with the sciences and some SDSs of the social sciences, for a total of 12,747 professors in 195 SDSs and 11 UDAs, as shown in Table 1.[3]

The bibliometric dataset used to measure output is extracted from the Observatory of Public Research (ORP), a database developed by the authors and derived under license from Clarivate Analytics' Web of Science (WoS). Beginning from the raw data of WoS and applying a complex algorithm for disambiguation of the true identity of the authors and reconciliation of their institutional affiliations, each publication is attributed to the university professor that produced it, with a harmonic average of precision and recall (F-measure) equal to 97%.

---

[1] The complete list is accessible at http://attiministeriali.miur.it/UserFiles/115.htm, last accessed October 25, 2018.
[2] http://cercauniversita.cineca.it/php5/docenti/cerca.php, last accessed October 25, 2018.
[3] The analysis omits the SDSs where over 50% of professors have no publications indexed in the WoS, over the period of observation.

*Table 1: Dataset of the anaylis by UDA*

| UDA | No. of SDSs | No. of Italian professors in the dataset |
|---|---|---|
| 1 - Mathematics and computer science | 9 | 1,199 |
| 2 - Physics | 8 | 1,074 |
| 3 - Chemistry | 11 | 1,361 |
| 4 - Earth sciences | 12 | 407 |
| 5 - Biology | 19 | 1,983 |
| 6 - Medicine | 47 | 3,528 |
| 7 - Agricultural and veterinary sciences | 29 | 909 |
| 8 - Civil engineering | 9 | 386 |
| 9 - Industrial and information engineering | 41 | 1,578 |
| 11 - Psychology | 5 | 103 |
| 13 - Economics and statistics | 5 | 219 |
| Total | 195 | 12,747 |

The identification of the TSs requires the measurement of research performance for all professors. The bibliometric indicator of performance used is Fractional Scientific Strength (FSS). The FSS is a proxy of average yearly total scholarly impact[4] of an individual's research activity over a period of time. Most bibliometricians define productivity as the number of publications in the period of observation. Because publications have different values (impact), we prefer to adopt a more meaningful definition of productivity: the value of output per unit value of labor, all other production factors being equal. The latter recognizes that the publications embedding new knowledge have a different value or impact on scientific advancement. At present we provide the formula to measure FSS, while referring the reader to Abramo and D'Angelo (2014) for a thorough treatment of the underlying microeconomic theory and all the limits and assumptions embedded in both the definition and the operationalization of the measurement.

$$FSS = \frac{1}{t}\sum_{i=1}^{N}\frac{c_i}{\bar{c}}f_i$$

[1]

Where:
$t$ = number of years of work in the period under observation
$N$ = number of publications in the period under observation
$c_i$ = citations received by publication $i$
$\bar{c}$ = average number of citations received for all cited publications in same year and subject category of publication $i$
$f_i$ = fractional contribution of professor to publication $i$.

The fractional contribution equals the inverse of the number of authors in those fields where the practice is to place the authors in simple alphabetical order, but assumes different weights in other cases. For the life sciences, widespread practice in Italy is for the authors to indicate the various contributions to the published research by order of names in the byline. For the life science SDSs, we therefore assign different weights to each coauthor according to their position in the list of authors and the character of the coauthorship (intra-mural or extra-mural).

---
[4] Refer to Abramo (2018) for a thorough discussion about the definition and bibliometric measurement of impact

The value of FSS is measured for all professors (including those who do not satisfy the conditions to belong to the dataset of analysis) in the SDSs under observation over two distinct periods: 2001-2005 and 2006-2010. Then for each period we identify the TSs as those placing among the top 10% by FSS, in each SDS.

In order to assess the collaboration behavior of professors in the dataset, we analyze the nature of co-authorships, adopting the taxonomy described in Abramo, D'Angelo, and Murgia (2013b): for each academic $i$ of the dataset, we measure the propensity to collaborate, both overall and for individual type of collaboration, through the following indicators:

- Propensity to collaborate: $C = \frac{cp_i}{N_i}$, where $cp_i$ is the number of publications resulting from collaborations (two or more co-authors in the byline) over the period and $N_i$ is the total number of publications written by the academic $i$ over the period;
- Propensity to collaborate at the intra-university level: $CI = \frac{cip_i}{N_i}$, where $cip_i$ is the number of publications resulting from collaborations with other academics belonging to the same university over the period;
- Propensity to collaborate extramurally at the domestic level: $CED = \frac{cedp_i}{N_i}$, where $cedp_i$ is the number of publications resulting from collaborations with scientists belonging to other domestic organizations over the period;
- Propensity to collaborate extramurally at the international level: $CEF = \frac{cefp_t}{N_i}$, where $cefp_i$ is the number of publications resulting from collaborations with scientists belonging to foreign organizations over the period.

These indicators vary between zero (if, in the observed period, the scientist under observation did not produce any publications resulting from the form of collaboration analyzed), and 1 (if the scientist produced all his/her publications through that form of collaboration).[5]

## 4. Results and analysis

Given the reciprocal influence between collaboration intensity and performance, the 2001-2010 period was broken into two five-year sub-periods, 2001-2005 and 2006-2010, and the bibliometric measures were calculated separately, as described above. The underlying rationale is that of verifying whether a change of status from TS to non-TS, or vice versa, corresponds to a change in the scientist's collaboration behavior.

Table 2 present the distribution of professors of the dataset, per UDA, divided into four classes:

---

[5] Similar indicators are presented by Martín-Sempere, Garzón-Garcia and Rey-Rocha (2008) and Ductor (2015).

- YES-YES: Professors classified as TS for FSS in both the first and second five-year period,
- YES-NO: professors who are TS in the first five-year period, but not the second,
- NO-YES: professors who are TS in the second period, but not the first,
- NO-NO: professors who are TS in neither period.

At the general level (last line) we observe that 91% of professors maintain the same status over both five-year periods; 4% improve their performance to the point of becoming TS in the second period, while 5% who were TS in the first period lose this status in the second.[6]

The percentage of those who maintain TS status over the entire 10 years varies between the 4% of UDA 9 and 15% of UDA 11. Comparison between columns 3 and 4 shows that in nine out of 11 UDAs there are greater percentages of professors experiencing a downgrade (YES-NO) than those experiencing an upgrade (NO-YES).

*Table 2: Distribution of the professors of the dataset into the four classes of belonging to TS (2001-2005 vs 2006-2010), per UDA*

| UDA* | NO-NO | NO-YES | YES-NO | YES-YES | Total |
|---|---|---|---|---|---|
| 1 | 1,005 (84%) | 62 (5%) | 67 (6%) | 65 (5%) | 1,199 |
| 2 | 931 (87%) | 33 (3%) | 50 (5%) | 60 (6%) | 1,074 |
| 3 | 1,139 (84%) | 52 (4%) | 66 (5%) | 104 (8%) | 1,361 |
| 4 | 337 (83%) | 21 (5%) | 27 (7%) | 22 (5%) | 407 |
| 5 | 1,671 (84%) | 83 (4%) | 79 (4%) | 150 (8%) | 1,983 |
| 6 | 2,956 (84%) | 124 (4%) | 180 (5%) | 268 (8%) | 3,528 |
| 7 | 743 (82%) | 45 (5%) | 63 (7%) | 58 (6%) | 909 |
| 8 | 306 (79%) | 17 (4%) | 32 (8%) | 31 (8%) | 386 |
| 9 | 1,346 (85%) | 73 (5%) | 92 (6%) | 67 (4%) | 1,578 |
| 11 | 79 (77%) | 6 (6%) | 3 (3%) | 15 (15%) | 103 |
| 13 | 163 (74%) | 13 (6%) | 23 (11%) | 20 (9%) | 219 |
| Total | 10,676 (84%) | 529 (4%) | 682 (5%) | 860 (7%) | 12,747 |

*\* 1 - Mathematics and computer science, 2 - Physics, 3 - Chemistry, 4 - Earth sciences, 5 - Biology, 6 - Medicine, 7 - Agricultural and veterinary sciences, 8 - Civil engineering, 9 - Industrial and information engineering, 11 - Psychology, 13 - Economics and statistics*

**4.1 Variation in the propensity to collaborate**

Table 3 presents the average value of propensity to collaborate (C) in the two five-year periods, in general and for all four classes considered (TS and non, 2001-2005 vs 2006-2010). Taking all 12,747 professors of the dataset, the propensity to collaborate varies from an average of 96.7% in the first period to 97.3% in the second, confirming all previous studies in literature indicating research as an increasingly collaborative activity. The professors with a positive change in value of C over the two periods are 92.4% of total; a minimum average of 82.4% is observed in the two classes NO-YES and YES-YES.

The difference between the averages varies between +0.3 percentage points (for the NO-YES class) to a maximum of +1.9 (in the YES-NO class). The *t*-test shows statistical significance for these differences. A summary reading of the data leads to the observation that the increase in propensity to collaborate is less for scientists who remain or become TS, than it is for their less-performing peers. In other words, the propensity to collaborate in scientific activities shows a growing trend, but this trend seems to only marginally affect top scientists.

---
[6] Note that the percentages of NO-YES TSs and YES-NO TSs do not coincide because TSs are the top productive professors in the overall population (larger than the dataset under analysis).

*Table 3: Professors' average propensity to collaborate (C) in the two five-year periods*

| TS class† | Obs | Average 2001-2005 | Average 2006-2010 | Δ | % professors with delta >=0 | t | |
|---|---|---|---|---|---|---|---|
| NO-NO | 9,673 | 96.9% | 97.4% | +0.5% | 94.0% | -4.05 | *** |
| NO-YES | 529 | 95.8% | 96.0% | +0.3% | 82.4% | -0.47 | |
| YES-NO | 671 | 95.2% | 97.0% | +1.9% | 90.5% | -3.70 | *** |
| YES-YES | 860 | 96.0% | 97.0% | +0.9% | 82.4% | -3.48 | *** |
| Total | 11,733 | 96.7% | 97.3% | +0.6% | 92.4% | -5.44 | *** |

†*TS 2001-2005 vs 2006-2010*
\* p <0.1; ** p < 0.05; *** p <0 .01
*Two-sample paired t-test with unequal variances; two-tailed - level (0.05)*

The same type of analysis was repeated for each type of collaboration (international, extramural domestic, and intramural).

Table 4 presents the results for propensity to collaborate at the international level (CEF). As for collaboration in general, the distributions for the two five-year periods show differences that are always statistically significant. In particular, the average for this type of collaboration behavior for all professors of the dataset rises from 21.0% for 2001-2005 to 23.8% for 2006-2010, in accordance with a world trend already observed in the literature. However the greatest increase in propensity to international collaboration (+5.6%) is seen for professors of the NO-YES class, and next for those who maintain their TS status (YES-YES, increase +3.7%). Although at a smaller pace (+2.5%), also professors who remain non TS (NO-NO class) show an increased propensity to collaborate. In summary it seems that the increase in propensity to collaborate at international level is greater for scientists who become or remain TS, than it is for others. The result may underline a bi-univocal relation between productivity and collaboration, whereby the status of TS attracts the attention of foreign colleagues and, at the same time, participation in international research projects help boost productivity.

*Table 4: Professors' average propensity to collaborate at international level (CEF) in the two five-year periods*

| TS class† | Average 2001-2005 | Average 2006-2010 | Δ | % professors with delta >=0 | t | |
|---|---|---|---|---|---|---|
| NO-NO | 19.9% | 22.4% | +2.5% | 71.2% | -9.11 | *** |
| NO-YES | 23.1% | 28.7% | +5.6% | 66.7% | -5.66 | *** |
| YES-NO | 25.2% | 28.1% | +2.9% | 62.9% | -3.56 | *** |
| YES-YES | 29.0% | 32.7% | +3.7% | 63.4% | -6.42 | *** |
| Total | 21.0% | 23.8% | +2.8% | 69.9% | -11.49 | *** |

†*TS 2001-2005 vs 2006-2010*
\* p <0.1; ** p < 0.05; *** p <0 .01

The analysis concerning the propensity for extramural domestic collaboration (CED) shows average differences between the two five-year periods consistently positive and statistically significant (Table 5). At the general level, the observations rise from an average value of 43.5% for 2001-2005 to 50.3% for 2006-2010 (delta +6.8%). Analyzing the individual classes, we observe that this increase largely concerns professors who lose the TS status (YES-NO, +7.9%) or who never had it (NO-NO, +7.0%). At the same time, the data in column 5 indicate that the remaining two classes (YES-YES and NO-YES) are those that register the lowest percentage of professors who increase their value of CED. In summary, the increase in propensity to collaborate at extramural domestic level is less for scientists who remain or become TS, compared to their colleagues in the other classes.

*Table 5: Professors' average propensity to collaborate at extramural domestic level (CED) in the two five-year periods*

| TS class† | Average 2001-2005 | Average 2006-2010 | Δ | % professors with delta >=0 | t | |
|---|---|---|---|---|---|---|
| NO-NO | 43.7% | 50.8% | 7.0% | 67.0% | -19.21 | *** |
| NO-YES | 41.9% | 45.7% | 3.8% | 59.0% | -3.36 | *** |
| YES-NO | 40.6% | 48.5% | 7.9% | 64.8% | -7.70 | *** |
| YES-YES | 44.2% | 49.7% | 5.4% | 62.9% | -7.96 | *** |
| Total | 43.5% | 50.3% | 6.8% | 66.2% | -21.60 | *** |

†*TS 2001-2005 vs 2006-2010*
\* $p <0.1$; \*\* $p < 0.05$; \*\*\* $p <0.01$

Finally, Table 6 presents the results concerning propensity to collaborate at intramural level (CI). For this case we again observe positive and statistically significant changes in collaboration propensities, between the two five-year periods for the different classes, with the exception of the "NO-YES" class. Although the professors of this class do on average increase their performance to the point of achieving TS status, on the other hand they also show a reduction (-0.4%) of propensity to collaborate with colleagues from the same university. On the contrary, the colleagues that lose top status (YES-NO class) show a significant increase in CI (+3.6%). In summary, the increase in propensity to collaborate at intramural level is greater for scientists who lose TS status, compared to those who acquire it.

*Table 6: Professors' average propensity to collaborate at intramural level (CI) in the two five-year periods*

| TS class† | Average 2001-2005 | Average 2006-2010 | Δ | % professors with delta >=0 | t | |
|---|---|---|---|---|---|---|
| NO-NO | 76.9% | 77.7% | 0.8% | 68.6% | -2.72 | *** |
| NO-YES | 70.0% | 69.6% | -0.4% | 56.1% | 0.34 | |
| YES-NO | 67.8% | 71.3% | 3.6% | 64.2% | -3.59 | *** |
| YES-YES | 66.8% | 69.4% | 2.6% | 59.9% | -4.08 | *** |
| Total | 75.3% | 76.3% | 1.0% | 67.2% | -4.01 | *** |

†*TS 2001-2005 vs 2006-2010*
\* $p <0.1$; \*\* $p < 0.05$; \*\*\* $p <0.01$

We summarize the results of the analysis in Figure 1, which plots the differences between the average propensities to collaborate in all various forms, recorded in the two five-year periods. It is shown that professors who become TSs (NO-YES) register on average the maximum increase in the propensity to collaborate with foreign colleagues, followed by those remaining TS (YES-YES). Furthermore, professors who loose the TS status (YES-NO) register on average the maximum increase in the propensity to collaborate intramurally (CI), extramurally domestic (CED), and in general (C), showing that research collaboration does not necessarily help increase productivity.

*Figure 1: Percentage variations in propensity to collaborate in the two five-year periods for the four TS classes of professors in the dataset*

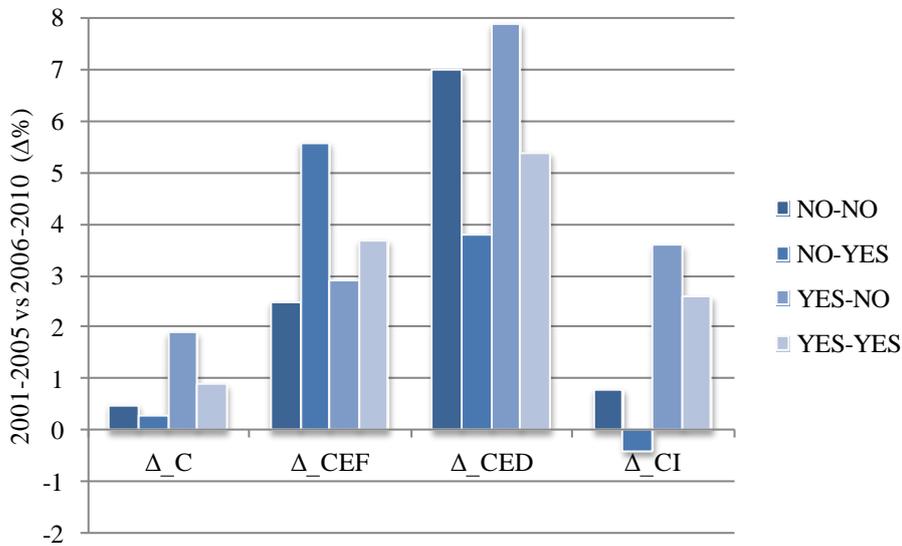

*C, Propensity to collaborate; CEF, Propensity to collaborate at international level; CED, Propensity to collaborate extramurally at the domestic level; CI, Propensity to collaborate at the intramural level*

### 4.2 Differences among disciplines

The analysis was repeated, dividing the population by UDA. Table 7 reports the average variation for each type of collaboration (2006-2010 vs 2001-2005), in terms of the different indicators, for the professors of the dataset divided by class and UDA.[7] The data confirm the observations at the general level, but with some interesting exceptions.

Those who are or become TS show no increase in their overall propensity to collaborate: the exceptions are in Mathematics and computer science, Civil engineering, and Industrial and information engineering. The propensity to collaborate at international level increases more for those who become or remain TS, in all disciplines except Physics and Chemistry. The propensity to collaborate at extramural domestic level is seen to increase less for those who become TS than for those who remain below "top", in all disciplines with the exceptions of Mathematics and computer science, and Industrial and information engineering. Similarly, the propensity for intramural collaboration increases less for those who become TS than for those who do not, in all disciplines but three: Earth sciences, Medicine, and Civil engineering.

---

[7] The analysis does not include UDAs 11 and 13 (Pedagogy and psychology; Economics and statistics), given the low number of observations in some classes of collaboration.

*Table 7: Average variation (2006-2010 vs 2001-2005) of the indicators of collaboration, for professors by UDA and classes of TS*

| UDA[†] |     | NO-NO |     | NO-YES |     | YES-NO |     | YES-YES |     |
|---|---|---|---|---|---|---|---|---|---|
| 1 | C   | 2.8%  | *** | 3.9%   |     | 7.6%   | *** | 6.3%   | *** |
|   | CEF | 1.8%  |     | 8.5%   | *** | 4.7%   | *** | 2.0%   |     |
|   | CED | 3.0%  | **  | 5.1%   | *   | 3.7%   | *** | 3.8%   | *   |
|   | CI  | 4.1%  | *** | -0.7%  |     | 5.9%   | *** | 2.1%   |     |
| 2 | C   | 0.4%  |     | -3.2%  | **  | 0.3%   | *** | 0.2%   |     |
|   | CEF | 4.1%  | *** | 0.0%   |     | -2.1%  | *** | 3.1%   |     |
|   | CED | 0.6%  |     | -6.0%  |     | -1.2%  | *** | -3.9%  | *   |
|   | CI  | 1.8%  | *   | -10.6% | **  | 6.2%   | *** | 6.2%   | **  |
| 3 | C   | 0.2%  |     | -2.5%  |     | 1.4%   | *** | 0.0%   |     |
|   | CEF | 3.8%  | *** | 5.0%   |     | 8.9%   | *** | 4.4%   | *** |
|   | CED | 4.5%  | *** | 2.4%   |     | 7.0%   | *** | 3.1%   | *   |
|   | CI  | -0.2% |     | -2.2%  |     | -0.3%  | *** | 1.2%   |     |
| 4 | C   | -0.4% |     | -1.7%  |     | 2.5%   | *** | 2.0%   |     |
|   | CEF | 3.4%  |     | 7.4%   |     | 4.5%   | *** | -1.0%  |     |
|   | CED | 1.7%  |     | -16.6% | **  | 13.2%  | *** | 8.3%   |     |
|   | CI  | -2.4% |     | 3.6%   |     | 9.6%   | *** | -5.9%  |     |
| 5 | C   | 0.0%  |     | 0.0%   |     | 0.6%   | *** | -0.2%  |     |
|   | CEF | 2.7%  | *** | 6.3%   | *** | 2.5%   | *** | 1.1%   |     |
|   | CED | 9.6%  | *** | 7.1%   | **  | 6.7%   | *** | 9.2%   | *** |
|   | CI  | 0.4%  |     | -1.8%  |     | 4.1%   | *** | 2.8%   | *   |
| 6 | C   | 0.1%  |     | -0.1%  |     | -0.2%  | *** | 0.2%   |     |
|   | CEF | 1.8%  | *** | 3.3%   |     | 1.3%   | *** | 3.4%   | *** |
|   | CED | 13.1% | *** | 9.4%   | *** | 14.7%  | *** | 9.4%   | *** |
|   | CI  | 0.3%  |     | 3.7%   |     | -0.5%  | *** | 2.5%   | **  |
| 7 | C   | 0.0%  |     | -0.7%  | *   | 2.6%   | *** | 0.2%   |     |
|   | CEF | 4.3%  | *** | 7.6%   | *   | 4.0%   | *** | 4.8%   | *   |
|   | CED | 8.6%  | *** | 0.2%   |     | 13.2%  | *** | 5.9%   | *   |
|   | CI  | -0.6% |     | -2.0%  |     | 3.0%   | *** | -0.1%  |     |
| 8 | C   | 0.3%  |     | 1.5%   |     | -0.4%  | *** | 1.3%   |     |
|   | CEF | 1.3%  |     | 7.7%   | *** | 0.3%   | *** | 7.2%   | *   |
|   | CED | 7.2%  | *** | -2.4%  |     | -1.0%  | *** | -4.5%  | *   |
|   | CI  | -0.4% |     | 1.7%   |     | 2.5%   | *** | 9.0%   | **  |
| 9 | C   | 0.3%  |     | 0.6%   |     | 2.0%   | *** | 1.4%   |     |
|   | CEF | 1.5%  | **  | 6.1%   | *** | 3.7%   | *** | 9.0%   | *** |
|   | CED | 1.5%  |     | 3.4%   |     | 0.6%   | *** | -1.2%  |     |
|   | CI  | 1.2%  |     | 0.5%   |     | 9.2%   | *** | 3.6%   |     |

[†] 1 - Mathematics and computer science, 2 - Physics, 3 - Chemistry, 4 - Earth sciences, 5 - Biology, 6 - Medicine, 7 - Agricultural and veterinary sciences, 8 - Civil engineering, 9 - Industrial and information engineering
\* p <0.1; \*\* p < 0.05; \*\*\* p <0 .01

### 4.3 International collaborations, impact and research performance

From the above analyses, it seems that the increase in propensity to international collaboration is typical of the NO-YES class, meaning for those who significantly increase their performance to the point of rising in the ranking of their SDS and reaching the first decile for FSS. This could be due to the greater average impact precisely of those publications achieved in collaboration with their foreign colleagues. To test this possibility, Figure 2 reports, for each UDA, the average values of field-normalized impact[8] of the publications achieved by the professors (subdivided between TSs and "others") in the two periods under consideration (see the Appendix for details).

For both of the five-year periods, the average impact of the publications achieved by the TSs is greater than that of the publications achieved by their peers, in all UDAs, both for "domestic" publications (with a byline consisting of only Italian authors) and for "international" ones. Also, the publications that are the result of international collaborations are on average more cited than the domestic ones, independent of the UDA or rank of author (TS or other). This confirms our hypothesis.

*Figure 2: Average field-normalized impact of publications by top scientists vs others, by UDA*

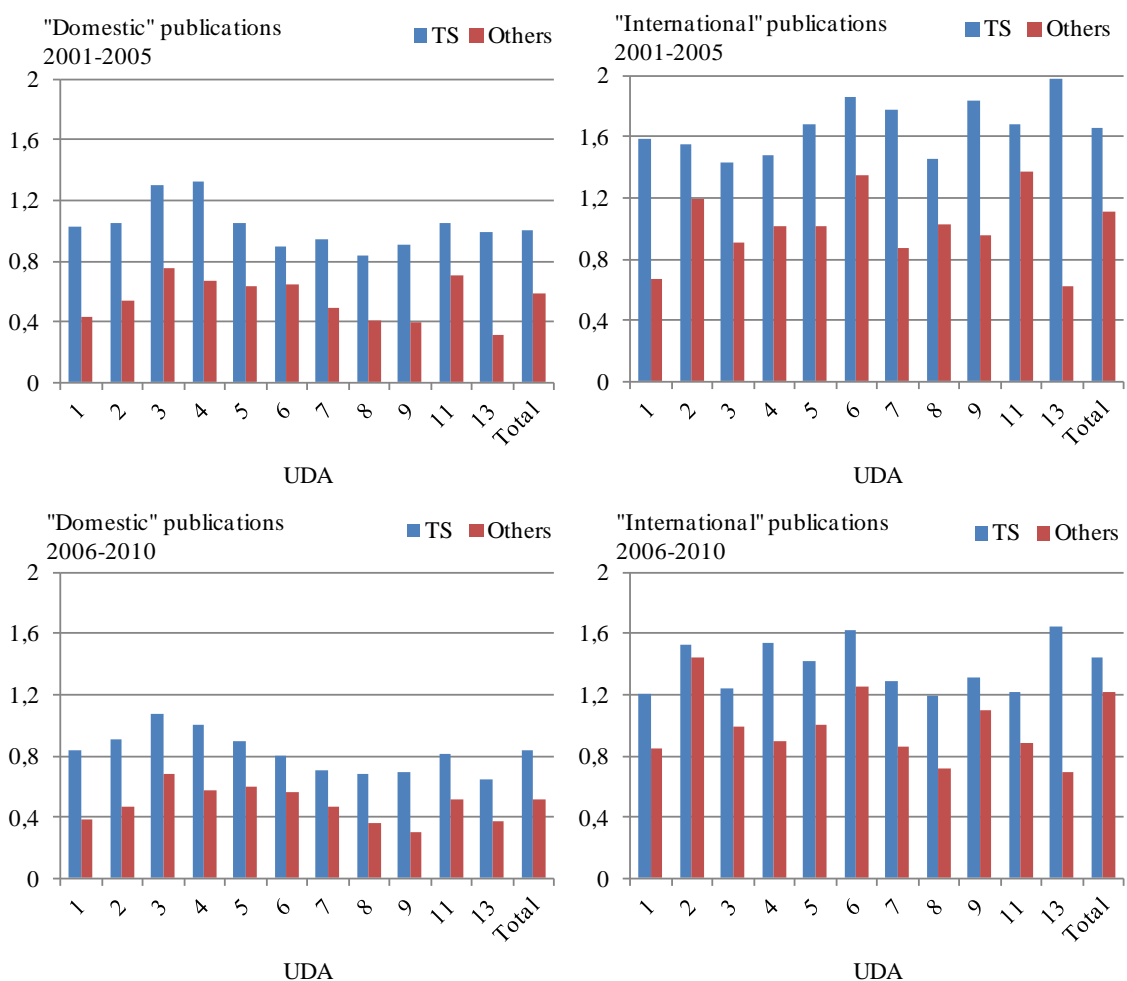

*UDA: 1 - Mathematics and computer science, 2 - Physics, 3 - Chemistry, 4 - Earth sciences, 5 - Biology, 6 - Medicine, 7 - Agricultural and veterinary sciences, 8 - Civil engineering, 9 - Industrial and information engineering, 11 - Psychology, 13 - Economics and statistics*

## 5. Conclusions

A growing number of governments place priority on continuous improvement in effectiveness and efficiency of their research systems.

---
[8] With reference to equation [1].

Collaboration plays a crucial role among the factors influencing research performance. The relationship between intensity of collaboration and research performance has been amply investigated in the literature, not always with convergent results. Both performance and research collaboration are difficult to define and measure, and this contributes to the challenge of clarifying the causal nexus. In fact collaboration influences research productivity (because it impacts on its critical factors – competencies, resources, time, etc.), but at the same time research productivity influences collaboration (thanks to the "attraction" exercised by the most productive scientists).

This work has explored the nexus between research collaboration and productivity, in particular by comparing the behavior of TSs of the different fields with that of their less productive colleagues. We have analyzed collaboration behaviors at the "international", "domestic extramural" and "intramural" levels, to understand in what manner these differentiate between the two groups, of TSs and non, in like fields.

The empirical analyses examine the coauthorship of scientific publications by more than 12,000 Italian professors, observed over two successive five-year periods. The results are interesting. On the one hand, the analysis registers a significant increase in the general propensity to collaborate, from an average of 96.7% in the first period to 97.3% in the second, which is in line with the literature (Abt, 2007; Uddin, Hossain, Abbasi, & Rasmussen, 2012). However the data show that the increase is less for scientists who remain or become TS than it is for the rest of the population.

The increase in the resort to collaboration concerns all three types of behavior, but is particularly notable in regards to domestic extramural collaboration, which in the observed population increases from an average of 43.5% in 2001-2005 to 50.3% in 2006-2010. Still, this increase is less for scientists who remain or become TS than it is for their colleagues of other classes.

The latter is true also with regard to intramural collaborations. In fact here, scientists who increase their productivity to the point of becoming TS actually register a decrease in intramural collaborations, of -0.4%.

The propensity to collaborate at international level once again increases at the general level (delta +2.8% for average values), from one five-year period to the next. However in contrast, concerning this behavior, those scientists who become or remain TS register significantly greater increases (+5.6%, +3.7%) than do their colleagues. This finding is aligned with Abramo et al. (2011).

The increase in productivity by those who acquire TS status seems due precisely to the greater average impact of the publications achieved in collaboration with foreign colleagues. The analyses repeated by discipline confirm the observations at the general level, with a few exceptions.

What the results show is that scientists who move to top tend to decrease intramural collaborations in favor of a relatively significant increase in international ones, which lead to publications with higher average impact.

A greater impact of publications coauthored with foreign colleagues may be due to different factors. The content and scope of the publication may be more international, attracting then the attention of a wider audience and therefore being more likely cited. Also the quality of publications may benefit from an international research team, who bring different resources and perspectives. Finally, the quality of authors engaging in international collaborations should be above standards in the first place, in order to attract or be chosen by foreign collaborators.

From a policy perspective, because, all others equal, increase in productivity is the underlying aim of all productive systems, fostering international collaboration is an indirect way to achieve it. To foster the propensity to collaborate at the international level, a wide variety of incentives can be envisaged. Increasing the freedom and responsibility of individual researchers and research organizations to form international research partnerships and attract foreign researchers. Utilizing honorary and visiting professor or research-fellow appointments to attract external scholars for collaboration purposes. The creation of internationalization offices, focused on promotion of the institutions research qualities and strengths. Finally, funding schemes can be specifically engineered to require partnerships among individuals or organizations, thus facilitating bottom-up collaboration.

The usual warnings on the generalization of results to other countries apply. As Aldieri et al. (2017) showed, the impact of internal and external collaborations on research performance is sensitive to the geographical dimension of the data.

*APPENDIX - Average field-normalized impact of publications of the professors in the dataset, by UDA*

|  | 2001-2005 | | | | | | 2006-2010 | | | | | |
|---|---|---|---|---|---|---|---|---|---|---|---|---|
|  | "domestic" publication | | | "international" publication | | | "domestic" publication | | | "international" publication | | |
| UDA* | TS | Others | Total | TS | Others | Total | TS | Others | total | TS | Others | total |
| 1 | 1.030 | 0.437 | 0.584 | 1.591 | 0.674 | 0.963 | 0.837 | 0.382 | 0.505 | 1.211 | 0.853 | 0.966 |
| 2 | 1.046 | 0.534 | 0.671 | 1.548 | 1.198 | 1.258 | 0.906 | 0.474 | 0.582 | 1.528 | 1.444 | 1.461 |
| 3 | 1.298 | 0.754 | 0.903 | 1.430 | 0.907 | 1.121 | 1.081 | 0.680 | 0.800 | 1.238 | 0.988 | 1.082 |
| 4 | 1.329 | 0.674 | 0.848 | 1.474 | 1.010 | 1.201 | 1.009 | 0.573 | 0.665 | 1.543 | 0.900 | 1.075 |
| 5 | 1.051 | 0.631 | 0.754 | 1.685 | 1.021 | 1.267 | 0.897 | 0.595 | 0.691 | 1.423 | 1.005 | 1.165 |
| 6 | 0.900 | 0.652 | 0.744 | 1.866 | 1.351 | 1.598 | 0.799 | 0.561 | 0.647 | 1.620 | 1.253 | 1.419 |
| 7 | 0.945 | 0.493 | 0.627 | 1.783 | 0.877 | 1.275 | 0.705 | 0.464 | 0.533 | 1.292 | 0.864 | 1.033 |
| 8 | 0.840 | 0.412 | 0.595 | 1.450 | 1.025 | 1.258 | 0.678 | 0.359 | 0.500 | 1.196 | 0.720 | 0.999 |
| 9 | 0.907 | 0.395 | 0.520 | 1.839 | 0.961 | 1.231 | 0.692 | 0.306 | 0.402 | 1.313 | 1.101 | 1.176 |
| 11 | 1.046 | 0.712 | 0.832 | 1.688 | 1.369 | 1.511 | 0.812 | 0.521 | 0.654 | 1.220 | 0.885 | 1.122 |
| 13 | 0.992 | 0.316 | 0.538 | 1.976 | 0.626 | 1.377 | 0.651 | 0.367 | 0.462 | 1.644 | 0.694 | 1.261 |
| Total | 1.003 | 0.592 | 0.718 | 1.660 | 1.113 | 1.287 | 0.842 | 0.517 | 0.618 | 1.446 | 1.218 | 1.292 |

*\* 1 - Mathematics and computer science, 2 - Physics, 3 - Chemistry, 4 - Earth sciences, 5 - Biology, 6 - Medicine, 7 - Agricultural and veterinary sciences, 8 - Civil engineering, 9 - Industrial and information engineering, 11 - Psychology, 13 - Economics and statistics*